\documentclass[12pt]{article}
\def\theequation{\arabic{equation}}
\def\reset{\setcounter{equation}{0}}

\def\beq {\begin{eqnarray}}
\def\eeq {\end{eqnarray}}
\def\be {\begin{equation}}
\def\ee {\end{equation}}
\def \half {{\textstyle{1\over 2}}}
\def\vf {{\varphi}}
\def\Tr {{\rm Tr}}
\def\dag {\dagger}
\def\del {\partial}
\def\bdel{\bar{\partial}}

\def\e {\epsilon}
\def\d {\delta}
\def\bz {\bar{z}}
\def\half {{\textstyle {1 \over 2}}}
\def\vf {\varphi}
\def\ra {\rangle}
\def\la {\langle}
\def\Tr {{\rm Tr}}
\def\bA {{\bar A}}
\def\bD {{\bar D}}
\def\bV {{\bar V}}
\def\bz {{\bar z}}

\def \C {{\cal C}}

\def \l {{\lambda}}
\def \S {I}
\def \vx {\vec{x}}
\def \vy {\vec{y}}
\def \vv {\vec{v}}
\def \vu {\vec{u}}
\def \vk {\vec{k}}

\def \bn {{\bar n}}
\def \ba {{\bar a}}
\def \s {{\sigma}}
\def \bS {{\bar S}}
\def \H {{\cal H}}
\def \G {{\cal G}}
\def \A {\bar{\cal A}}
\def \E {{\cal E}}
\def \O {{\cal O}}
\def \bk {\bar{k}}
\begin{document}

\begin{titlepage}
\null\vspace{-62pt}

\pagestyle{empty}
\begin{center}
\rightline{CCNY-HEP-00/4}
\rightline{KIAS-P00047}
\rightline{RU-00-6B}

\vspace{1.0truein} {\large\bf Manifest covariance and the
Hamiltonian approach to mass gap}
\vspace{.2in} {\large\bf in (2+1)-dimensional Yang-Mills theory}\\
\vspace{.5in}  { DIMITRA KARABALI $^{a,d}$, CHANJU KIM $^{b}$ and
V.P. NAIR $^{c,d}$
\footnote{{\it e-mail addresses:} karabali@fabbro.rockefeller.edu,
cjkim@kias.re.kr, vpn@ajanta.sci.ccny.cuny.edu} }\\
\vspace{.3in}  {\it $^a$ Department of Physics and Astronomy,
Lehman College of the CUNY\\  Bronx, NY 10468}\\
\vspace{.1in}  {\it $^b$ Korea Institute for Advanced Study\\
130-012 Seoul, South Korea}\\
\vspace {.1in}  {\it $^c$ Physics Department,   City College of the
CUNY\\  New York, NY 10031}\\
\vspace {.1in} {\it $^d$ Physics Department, Rockefeller
University\\ New York, NY 10021}\\
\vspace{0.5in}
\end{center}
\vspace{0.5in}

\centerline{\bf Abstract}

In earlier work we have given a Hamiltonian analysis of Yang-Mills
theory in (2+1) dimensions showing how a mass gap could arise. In
this paper, generalizing and covariantizing from the mass term in
the Hamiltonian analysis, we obtain two manifestly covariant and
gauge-invariant mass terms which can be used in  a resummation of
standard perturbation theory to study properties of the mass gap.

\baselineskip=18pt

\end{titlepage}

\hoffset=0in
\newpage
\pagestyle{plain}
\setcounter{page}{2}
\newpage

\noindent{\bf 1. Introduction}
\vskip .1in In a series of recent papers we have carried out a
Hamiltonian analysis of Yang-Mills theories in (2+1) dimensions,
$YM_{2+1}$
\cite{KN, KKN1, KKN2}.  A matrix parametrization of the gauge
potentials
$A_{\mu}$ was used which facilitated calculations using manifestly
gauge-invariant variables. An analytical formula for the string
tension was obtained which was found to be in good agreement with
lattice gauge theory simulations \cite{KKN2, teper}.  It was also
shown that effectively the gauge bosons become massive. This mass
can be identified in the context of a (3+1)-dimensional gluon
plasma as the magnetic mass \cite{VPN1}. The analytically
calculated value of this mass is also in reasonable agreement with
numerical estimates \cite{karsch}.

All the above calculations were done in a Hamiltonian framework.
The virtue of this approach is that at a given time we have to
consider gauge potentials on the two-dimensional space and for
two-dimensional gauge fields a number of calculations can be done
exactly. However, as in any Hamiltonian analysis, we do not have
manifest Lorentz covariance. Overall Lorentz covariance is not
lost since the requisite commutation properties on the components
of the energy-momentum tensor may be verified \cite{KN}.   Now, the
main physical context in which our results could be applied would
be the case of magnetic screening in QCD at high temperatures. The
Wick-rotated version of $YM_{2+1}$, namely three-dimensional
Euclidean Yang-Mills theory, is what is needed to describe the
zero Matsubara frequency mode of the (3+1)-dimensional QCD at high
temperatures. A manifestly covariant formulation of our analysis
would be just what is ideal in relating our results to Feynman
diagrams in high temperature QCD.  There are two sources of lack
of manifest covariance in our approach, firstly  due to the use of
the Hamiltonian analysis itself and secondly, because the
gauge-invariant variables we used were defined intrinsically in a
(2+1)-splitting and do not have simple (tensorial) transformation
properties under Lorentz transformations. Going over to a
Lagrangian might address the first problem of degrees of freedom
being defined at a constant time but not the second, unless we
have  a Lorentz covariant parametrization of the gauge potentials
which makes it easy to isolate the gauge-invariant degrees of
freedom. In our approach, calculability, viz., the fact that the
transformation of variables could be done exactly, including the
Jacobian, was the crucial factor, which led to physical results.
To be useful to a similar degree, one needs a Lorentz covariant
parametrization of $A_{\mu}$ for which the change of variables to
the gauge-invariant degrees of freedom can be carried out,
including the path-integral Jacobian in a nonperturbative way. We
have not been able to find such a set of variables. The situation
is similar to the old problem of rewriting Yang-Mills theory in
terms of Wilson loop variables and other similar choices of
variables; as in many earlier attempts, the technical stumbling
block has been the calculation of Jacobians in nonperturbative
terms.

A more practical alternative strategy would then be the following.
First of all, we can consider an expansion of our results in
powers of the coupling constant. It then becomes clear that the
mass gap cannot be seen to any finite order in the perturbative
expansion but could be obtained by resummation of certain series
of terms. Such resummations can be carried out in the covariant
path-integral approach by  adding and subtracting suitable
(gauge-invariant) mass terms, and indeed, many such calculations
have already been done using different choices of mass terms
\cite{AN, BP,JP}.  In these calculations, there is no unique or
preferred mass term we can use. The natural question is whether
our  Hamiltonian analysis can shed any light on this issue; in
other words, are there any mass terms which are similar or close
to the mass term which arises in the Hamiltonian  analysis?

In this paper we do the following. We study the properties of the
mass term which arises in our Hamiltonian analysis, identifying
certain key features and then seek covariant gauge-invariant mass
terms which can be used in a Lagrangian resummation procedure and
which are simple generalizations of what we find in the
Hamiltonian analysis. Two such terms are considered and analyzed
to some extent.

In the next section, we discuss an ``improved" version of
perturbation theory starting with our Hamiltonian analysis. We
first show how the mass term can be manifestly displayed to the
lowest order in our gauge-invariant variables. Then building upon
this lowest order result, we identify the required properties and
the nature of the mass term. A procedure for the covariantization
of the mass term is decribed in section 3.
 Explicit formulae for the covariantized mass terms are given to
cubic order in the potentials. Section 4 gives a brief discussion
of the results of carrying out the resummation to the lowest
nontrivial order.  The paper concludes with a short summarizing
discussion. Some technical arguments on the nature of the mass
term are given in the appendix.

\vskip .1in
\noindent{\bf 2. ``Improved" perturbation theory and the mass term}
\vskip .1in We consider an $SU(N)$-gauge theory with the gauge
potentials
$A_i = -i t^a A_i ^a$, $i=1,2$, where $t^a$ are hermitian
$(N \times N)$-matrices which form a basis of the Lie algebra of
$SU(N)$ with
$[t^a, t^b ] = i f^{abc} t^c,~~{\rm {Tr}} (t^at^b) = {1 \over 2}
\delta ^{ab}$.  The Hamiltonian analysis was carried out in the
$A_0=0$ gauge with the spatial components of the gauge potentials
parametrized as
\be A = -\partial M M^{-1},~~~~~~~~~~~~~~~~ \bar{A} = M^{\dagger
-1}
\bdel M^{\dagger}
\label{param}
\ee Here $A =\half (A_1 +i A_2), ~{\bar A} = \half (A_1 -i A_2),~
z=x_1 -ix_2,~{\bar z} =x_1+ix_2,~~ \del = \half
(\del_1+i\del_2),~\bdel =\half (\del_1-i\del_2)$. In the above
equation,  $M,~M^\dagger$ are complex
$SL(N,{\bf {C}})$-matrices. The volume element on the space
$\C$ of gauge-invariant configurations was calculated explicitly
in [1,2] and found to be
\be
 d\mu (\C )~= {[dAd\bA ]\over {\rm vol}{\cal G}}=~d\mu (H)
~e^{2c_A \S (H)}
 \label{measure}
 \ee where $H=M^{\dag} M$. $H$ is a gauge-invariant, hermitian
matrix-valued
 field.  $d\mu (H)$ is the Haar measure for $H$. (Explicitly, it
may be written as $ d\mu (H)  =  [d\vf ^a]\prod_{x} \det r $ where
$H^{-1}dH = d\vf ^a r_{ak} (\vf) t_k$.)
$c_A$ is the quadratic Casimir of the adjoint representation,
$c_A
\delta^{ab} = f^{amn}f^{bmn}$. $\S (H)$ is the  Wess-Zumino-Witten
(WZW) action for the hermitian matrix field
$H$ given by \cite{WZW1}
\be {\S} (H) = {1 \over {2 \pi}} \int \Tr (\partial H
\bar{\partial} H^{-1}) +{i
\over {12 \pi}} \int \epsilon ^{\mu \nu \alpha} \Tr ( H^{-1}
\partial _{\mu} H H^{-1}
\partial _{\nu}H H^{-1} \partial _{\alpha}H)
\label{WZW}
\ee As is typical for the WZW action, the second integral is over a
three-dimensional space whose boundary is the physical
two-dimensional space corresponding to the coordinates $z,~\bz$.
The integrand thus requires an extension of the matrix field $H$
into the interior of the three-dimensional space, but physical
results do not depend on how this extension is done
\cite{WZW1}. Actually for the special case of hermitian matrices,
the second term can also be written as an integral over spatial
coordinates only \cite{efraty}.

The inner product for two wavefunctions $\Psi_1,~\Psi_2$ is given
by
\be
\la 1 \vert 2\ra = \int d\mu (\C ) ~\Psi_1^* (H) \Psi_2 (H) =
\int  d\mu (H)  ~e^{2c_A \S (H)}~\Psi_1^* (H) \Psi_2 (H)
\label{inner}
\ee

Carrying out the change of variables from $A$ to $H$ in the
Hamiltonian operator, one gets
\beq
\H & = & T + V \nonumber\\ T & = & {e^2c_A\over 2\pi}\left[ \int_u
J^a(\vu) {\d \over \d J^a(\vu)} ~+~
\int
\Omega^{ab} (\vu,\vv)  {\d \over \d J^a(\vu) }{\d \over \d
J^b(\vv) }\right]
\nonumber\\
 V & = & {\pi \over {m c_A}} \int \bdel J_a \bdel J_a \\
 J & = & {c_A \over \pi} \del H H^{-1} \nonumber \\
 \Omega^{ab}(\vu,\vv)& = & {c_A\over \pi^2} {\d^{ab} \over
(u-v)^2} ~-~
 i {{f^{abc} J^c (\vv)}\over {\pi (u-v)}}\nonumber
\eeq

The first term in the kinetic energy $T$, viz., $J(\delta /\delta
J)$ shows that every power of $J$ in the wavefunction will give a
contribution $m= e^2 c_A / 2\pi$ to the energy. This is the basic
mass gap of the theory.

The volume element (\ref{measure}) plays a crucial role in how the
theory develops a mass gap. If
$\S (H)$ is expanded in powers of the magnetic field
$B^a= \half \epsilon_{ij} (\partial_i A_j^a - \partial_j A_i^a
+f^{abc}A_i^b A_j^c)$, the leading term has the form
\be
\S (H) \approx {1 \over 4\pi} \int B \left( {1 \over
\nabla^2}\right) B + {\cal O} (B^3)
\label{expansion}
\ee   Writing $\Delta E, ~\Delta B$ for the root mean square
fluctuations of the   electric field $E$  and the magnetic field
$B$, we have, from the canonical commutation rules
$[E_i^a, A_j^b]= -i\delta_{ij}\delta^{ab}$, $\Delta E~\Delta B\sim
k$,   where $k$ is the  momentum variable. This gives an estimate
for the energy
\be  {\E}={1\over 2} \left( {e^2 k^2\over\Delta B^2 } +{\Delta B^2
\over e^2}
 \right)
\ee  For low lying states, we must minimize ${\E}$ with respect to
$\Delta B^2$,
$\Delta B^2_{min}\sim  e^2 k$, giving ${\E}\sim k$. This
corresponds to the standard photon.  For the nonabelian theory,
this is inadequate since $\la \H \ra$  involves  the factor
$e^{2c_A \S (H)}$. In fact,
\be
\la \H \ra ~= \int d\mu (H)  ~e^{2c_A \S (H)}~ \half (e^2 E^2
+B^2/e^2 )
\ee  Equation (\ref{expansion}) shows that $B$ follows a Gaussian
distribution of width $\Delta B^2 \approx \pi k^2 /c_A$ for
small values of $k$. This Gaussian dominates near small $k$
giving $\Delta B^2 \sim k^2 (\pi /c_A )$.   In other words, even
though ${\E}$ is minimized around $\Delta B^2 \sim k$,
probability is   concentrated around $\Delta B^2 \sim k^2 (\pi
/c_A)$. For the expectation  value of the energy,  we then find
${\E}\sim e^2c_A/2\pi  +{\O}(k^2)$. Thus the kinetic term in
combination with   the measure factor $e^{2c_A\S (H)}$ could lead
to a mass gap of order $e^2c_A$.   The argument is not rigorous,
but captures the essence of how a mass gap   arises in our
formalism \cite{KN}.

All we have done so far is to rewrite the theory in terms of
gauge-invariant variables without making any other approximation.
It is therefore possible to look at perturbation theory in this
version. Since
$c_A$ is quadratic in the structure constants
$f^{abc}$, the exponent in (\ref{measure}) would be considered a
second order effect in the perturbative expansion. The exponential
in (\ref{measure}) would be expanded in powers of $c_A$ and we
would not see a Gaussian distribution for the magnetic
fluctuations (of width $\sim k^2$). Hence the effect considered
above cannot be seen to any finite order.  The basic question we
are asking in this paper is whether one can incorporate the
effects of the nontrivial measure (\ref{measure}) and the
resultant mass term in a covariant path integral for diagrammatic
analysis. It is clear that this cannot be done at any finite order
in perturbation theory. However, one can define an ``improved"
perturbation theory where a partial resummation of the
perturbative expansion has been carried out \cite{KKN1}.  This
improvement would be equivalent to keeping the leading term of
$\S (H)$ as in (\ref{expansion}) in the exponent in
(\ref{measure}). For example, if we write
$H= e^{t^a\vf^a} \approx 1+t^a \vf^a$, as would be appropriate in
perturbation theory, we find
\be  d\mu (\C ) \simeq [d \vf ] e^{- {c_A \over 2\pi} \int \del
\vf ^a \bdel \vf ^a}~\left(1~+{\cal O}(\vf^3 )\right)\label{6}
\ee Correspondingly, $J^a\simeq {c_A\over\pi}\del\vf ^a$, and the
Hamiltonian has the expansion
\be
\H \simeq m \left[ \int \vf _a {\d \over {\d \vf _a}} + {\pi \over
c_A} \int
\Omega (\vx,\vy) {\d \over {\d \vf _a (\vx)}} {\d \over {\d \vf _a
(\vy)}} \right] + {c_A \over m\pi} \int \del
\vf _a (-\del
\bdel) \bdel \vf _a~+{\cal O}(\vf^3 )
\label{Ham}
\ee where $m = e^2 c_A /2 \pi$ and $\Omega (\vx, \vy) = - \int
{d^2 k \over (2\pi)^2} ~{e^{i k \cdot (x-y)}/ k\bk}$. The term
$\int\vf _a {\d / {\d \vf _a}}$ shows that every $\vf$ in a
wavefunction  would get a contribution
$m$ to the energy; this is essentially the mass gap again.

The mass term can also be written in a different way as follows.
We can absorb the exponential factor of (\ref{6}) into the
wavefunctions,  defining
$\Phi = e^{-{c_A \over 4\pi} \int
\del \vf \bdel
\vf} \Psi$, so that the norm of $\Phi$ 's involves just
integration of $\Phi ^*
\Phi$ with the flat measure $[d \vf ]$, i.e.,
\beq
\la 1 \vert 2\ra \approx \int  [d \vf ] ~~\Phi_1^* (H) \Phi_2
(H)\label{10'}
\eeq For the wavefunctions $\Phi$, we get
\beq
\H ' &&\simeq {1\over 2} \int_x \left[- {\d ^2 \over {\d \phi _a
^2 (\vx)}} + \phi _a (\vx)
\bigl( m^2 -
\nabla ^2 \bigr)  \phi _a (\vx)\right] + ...\nonumber\\ &&\simeq
{1\over 2} \int_x \left[- {\d ^2 \over {\d \phi _a ^2 (\vx)}} +
\phi _a (\vx)
\bigl( -
\nabla ^2 \bigr)  \phi _a (\vx) + {e^2c_A^2 \over 4\pi^2}~\partial
\vf_a \bdel \vf_a
\right]\label{10}
\eeq where $\phi _a (\vk) = \sqrt {{c_A k \bar{k} }/ (2 \pi m)}~
\vf _a (\vk)$.  Expression (\ref{10}) is the Hamiltonian for a
field of mass $m= e^2 c_A / 2\pi$. This can be taken as  the
lowest order term of an ``improved" perturbation theory. In the
second line of (\ref{10}), we have also separately shown the mass
term since we shall need it shortly.

It may be worth emphasizing that this Hamiltonian (\ref{10}), with
the inner product (\ref{10'}), is entirely equivalent to the
previous one (\ref{Ham}), with the inner product given by
(\ref{6}) \cite{sakita}. However, in (\ref{10}), the mass term has a more
conventional form and therefore one can use this as a starting
point for the mass term we want to find for  resummation
calculations in the Lagrangian formalism. We also see that the
energy of the particle, viz., $\sqrt{k^2 + m^2}$ is an infinite
series when expanded in powers of
$e^2$. The ``improved" perturbation theory, which is effectively
resumming this up, is thus equivalent to a partial resummation of
the perturbative expansion.

The gauge-invariant variables $\vf_a$ or $H$ are wonderfully
appropriate for the Hamiltonian analysis. However, in a
perturbative diagrammatic calculation carried out in a covariant
Lagrangian framework, we shall need to use the gauge potentials
$A_i$. To the lowest order, the number of powers of $\vf$'s and
$A_i$'s do match; the mass term given in (\ref{10}) is thus
quadratic in the $A$'s and can be written as
\be {e^2c_A^2 \over 4\pi^2}\int \partial \vf_a \bdel \vf_a
={m^2\over e^2}
\int {d^2k\over (2\pi )^2} A^a_i (-k) \left( \delta_{ij} - {k_ik_j
\over k^2}\right) A^a_j(k)\label{c1}
\ee This gives the mass term only to the quadratic order and does
not have the full nonabelian gauge  invariance; there will be
terms with higher powers of $A$'s giving a gauge-invariant
completion of (\ref{c1}). Already at this stage we can say
something about how the  full mass term should look like, based on
the following conditions.

\noindent 1. The mass term $F$ should be expressible in terms of
$H$ since that is the basic gauge-invariant variable of the
theory. (The $\vf_a$'s represent a particular way to parametrize
$H$. It should be possible to write the mass term in a way that is
not sensitive to how we parametrize $H$.)

\noindent 2. To the lowest, viz., quadratic order, it should agree
with the mass term in (\ref{10}) or (\ref{c1}).

\noindent 3. The mass term should have ``holomorphic invariance".

The last property is the following requirement. As can be seen
from the definitions (\ref{param}), the matrices $(M,~M^{\dag})$
and $(M\bV(\bz),~ V(z) M^{\dag})$ both define the same potentials
$(A, \bA)$, where $V(z)$ is holomorphic in $z$ and $\bV(\bz)$ is
antiholomorphic. In terms of $H$, this means that $H$ and $V H\bV$
are physically equivalent. Physical quantities should be, and in
any correct calculation will be, invariant under
$H\rightarrow VH\bV$, so that the ambiguity in the choice of the
matrices
$M,~M^\dag$ does not affect the physics. For example, the WZW
action in (\ref{WZW}) is invariant under $H\rightarrow VH\bV$, a
property familiar from two-dimensional physics. We have previously
referred to this invariance requirement as  ``holomorphic
invariance"; it can be used as a guide in some calculations.

A minimal mass term with the above requirements can be easily
written down. First of all, since $H= e^{t^a\vf^a}$, we see that,
in terms of $H$, the mass term shown in (\ref{10}) is of the form
$\Tr (\partial H \bdel H^{-1})$. (We shall discuss this in the
appendix in some detail. The key point is that we have $\Tr
(\partial H \bdel H^{-1})$ and not something like
$\Tr (\partial H \bdel H)$, eventhough the latter does have the
same kind of quadratic approximation.) Notice that this term, $\Tr
(\partial H \bdel H^{-1})$, is the first term of the WZW action
(\ref{WZW}). Since the WZW action has holomorphic invariance, we
see that a minimal mass term, or a minimal holomorphically
invariant completion of  $\Tr (\partial H \bdel H^{-1})$, with the
properties 1-3 listed above is also a WZW action, i.e.,
\be F_{min}= -{2 \pi \over e^2} \S (H)
\label{a18}
\ee Of course, one can always add gauge-invariant terms which
start with cubic or higher powers of $A$, which do not spoil the
requirement that it agrees with (\ref{10}) at the quadratic order.
In this sense the WZW action is only a minimal mass term, not
unique. (The quadratic part is, of course, unique.)

There are also other invariant ways to complete $\Tr (\partial H
\bdel H^{-1})$.  For example, we can write
\be F= {\pi^2 \over 2e^2c_A^2}\int ({\bar G}\bdel J^a )H_{ab} (G
\partial {\bar J}^b)
\label{a15}
\ee where $J^a =(c_A /\pi)(\partial H H^{-1})^a$,
$ {\bar J}^a = (c_A /\pi)(H^{-1}\bdel H)^a$ and $H_{ab} = 2 \Tr
(t_a H t_b H^{-1})$. This will  be holomorphically invariant with
the Green's functions
$G=\partial^{-1}$ and ${\bar G}= \bdel^{-1}$ transforming in a
certain way as discussed in \cite{KKN1}. This way of writing $F$
involves the additional use of Green's functions, over and above
the Green's functions which appear in the construction of $H$ (or
$M,~M^\dag$) from the potentials. In the next section, we give
expressions for the covariantized versions of both $F_{min}$ of
(\ref{a18}) and $F$ as in (\ref{a15}), to cubic order in
potentials. We shall see that $F$ equals $F_{min}$ plus a number
of terms which  involve the logarithms of momenta, the latter
having to do with the additional Green's functions. {\it The only
holomorphically invariant completion of
$\Tr (\partial H \bdel H^{-1})$ using $H$ and its derivatives, but
no additional Green's functions, is
$I(H)$ as given in (\ref{a18}).} This is why we refer to it as the
minimal term.

A strategy of doing the resummation calculations is then to
consider the action
\be S= S_{YM} - \mu^2 ~F_{min} ~+\Delta ~F_{min}
\label{19}
\ee  where we consider $\Delta$ to be of one higher order in a
loop expansion compared to $\mu^2$ and $S_{YM}$ is the usual
action for the YM path integral. In other words, the loop
expansion is organized by treating
$S_{YM} -\mu^2 F_{min}$ as the zeroth order term,  while $\Delta
F_{min}$ contributes at one loop higher. In particular
$\Delta$ is a parameter which is taken to have a loop expansion,
viz.,
$\Delta = \Delta^{(1)} + \Delta^{(2)} +...$. Since the parameter
$\mu^2$ is still arbitrary, we can choose it to be the exact value
of the pole of the full propagator. In other words, the pole of
the propagator (for the transverse potentials) remains
$\mu^2$ as loop corrections are added. This requires choosing
$\Delta^{(1)}$ to cancel the one-loop shift of the pole,
$\Delta^{(2)}$ to cancel the two-loop shift of the pole, etc.
$\Delta^{(1)},~\Delta^{(2)},$ etc. are calculated as functions of
the parameter $\mu^2$. The condition
$\mu^2 =\Delta$ then becomes a nonlinear equation for
$\mu^2$; it is the gap equation given as
\be
\Delta (\mu ) = \Delta^{(1)} + \Delta^{(2)} +... = \mu^2
\ee  This determines $\mu$ to the order to which the calculation
is performed. Thus in the end we also have $\mu^2 = \Delta$ as
desired. One can do similar resummation and gap equations wth any
mass term, for example $F$ in place of $F_{min}$ in (\ref{19}).

This procedure is, of course, what is done in any kind of
resummation or gap equation approach to mass generation \cite{AN,
BP, JP}. The additional ingredient for us is  that the Hamiltonian
analysis suggests some specific forms of the mass term
(\ref{a18}). The mass terms (\ref{a18}, \ref{a15}) are not
covariant, so we have to write covariantized versions of these
before they can be used in a  covariant resummation calculation.
We shall now consider a procedure for covariantization, which is
of interest in its own right.

\vskip .1in
\noindent{\bf 3. Covariantization of the mass term}
\vskip .1in
\noindent $\underline{General ~procedure}$
\vskip .1in

There is one more problem we have to deal with in using $\S (H) =
\S (A,~\bA)$ in a resummed perturbation theory, namely, that it is
not manifestly covariant. Again, the original theory is Lorentz
invariant and adding and subtracting $\S$ does not affect this.
However when we take $\mu^2$ and $\Delta$ to be of different
orders, we lose covariance order-by-order unless we use a
covariantized version of $\S$. In this section we outline a
general method of covariantization which can be used for $F$,
$F_{min}$.  Our method may also be interesting in its own right.

The key expressions we have involve holomorphic and
antiholomorphic derivatives and fields.
 We observe that
$\del = \half n_o^a \del_a$ and $\bdel = \half \bn_o^a \del_a$,
where $n_o^a = (1,i,0)$ and $\bn _o^a = (1,-i,0)$. Similarly for
the gauge fields $A,~\bA$.  An arbitrary Lorentz transformation of
$n_o^a$ and $\bn _o^a$ produces null 3-vectors $n^a,~ \bn^a$
respectively, such that
\beq  && n^a n_a = g_{ab} n^a n^b = 0 \cr && \bn^a \bn_a =  g_{ab}
\bn^a \bn^b = 0
\cr   && n^a \bn_a = g_{ab} n^a \bn^b =2
\label{null}
\eeq   where $g_{ab}$ is the Minkowski metric. We shall consider
the signature (1,1,-1). This suggests the following
covariantization procedure. Replace the holomorphic and
antiholomorphic derivatives ($\del ,~
\bdel$) and gauge fields ($A$,~$\bA$) in $I(H)$, expressed in
terms of the potentials, by ($\half n \cdot \del$, ~$\half \bn
\cdot \del$) and ($\half n \cdot A$,
$~\half \bn \cdot A$)  respectively, and then integrate over
Lorentz transformations. Thus the covariant analogue of a general
term
\be   S= \int dt d^2 x ~{\cal L} (A, \bA, \del, \bdel)
\ee   would be
\be   S_{cov} =\int d\mu \int dt d^2 x ~{\cal L} (\half {n \cdot
A},~\half{\bn \cdot A},~
\half {n \cdot
\del},~ \half{\bn \cdot \del})
\ee   where $d\mu$ is the measure over Lorentz transformations.

A particular parametrization for $n,~\bn$ is given by
\beq
 n_a &=& ( \cosh\theta \cos\chi - i \sin\chi, ~\cosh\theta
\sin\chi +i
\cos\chi, ~\sinh\theta)
\cr  \bn_a &=& ( \cosh\theta \cos\chi + i \sin\chi, ~\cosh\theta
\sin\chi -i \cos\chi, ~\sinh\theta)
\eeq   In terms of this parametrization, $d\mu = d(\cosh\theta )
d\chi$, where
$\cosh\theta\in (0, \infty)$ and $\chi \in (0, 2\pi)$.

The problem with this procedure is the fact that the Lorentz group
is noncompact and integration over Lorentz transformations leads
to divergences. The degree of divergence depends on the number of
$n$'s and $\bn$'s in the integrand. In order then for the
covariantization procedure to be meaningful one needs to regulate
the integrals in a consistent way. As we show below, this can be
done by  replacing the integrals by traces of suitable $(M\times
M)$-matrices. The integrals are then regained in a large
$M$-limit. To define the regularization, notice first of all that
there is no such  problem in Euclidean three-dimensional space.
Integration over Lorentz transformations is replaced by
integration over rotation angles and is convergent. This has been
used before in constructing covariant mass terms in Euclidean
space \cite{VPN2, AN}. The Euclidean version of the null vectors
$n$ is
\beq n_i &=& (-\cos\theta \cos\chi -i\sin\chi,~-\cos\theta
\sin\chi+i\cos\chi,~\sin\theta)\cr
\bn_i &=& (-\cos\theta \cos\chi +i\sin\chi,~-\cos\theta
\sin\chi-i\cos\chi,~\sin\theta)
\eeq The measure of integration over the angles is
$d\Omega=\sin\theta d\theta d\chi$. The Euclidean vectors
$n,~{\bar n}$ obey the same properties (\ref{null}), but with the
Minkowski metric replaced by the Euclidean one. If this procedure
is used for the  minimal mass term
$F_{min}= -({2 \pi /e^2}) \S (H)$, the resulting covariant mass
term is precisely what was proposed some time ago in \cite{VPN2}
and used in \cite{AN}. It is  interesting that this mass term
emerges in some minimal way  from our Hamiltonian analysis.

The Euclidean analysis is adequate for diagrammatic calculations.
However, conceptually, there is still something lacking.
Hamiltonian analysis is all in Minkowski space and to tie in
everything, it is important to define the covariantization
directly in Minkowski  space as well. In view of the Euclidean
result, one way to define the regularization of the integration
over the Lorentz transformations is then as follows. We do a Wick
rotation of the integrands to Euclidean space, do the integrals
there and then continue the final results back to Minkowski space.
Alternatively, one can seek a definition of the regularized
integrals in Minkowski space directly in such a way that the
results agree with the Wick rotation of the Euclidean results. We
now show how this can be done.

First we construct the operator analogues of the Minkowski null
vectors $n^a,~\bn^a$. Let
$g$ be a group element of $SO(2,1)$. $g$ can be written as $g =
e^{ i t^a
\theta^a}$ where
\be   t^a = (i\s_1, i\s_2, \s_3)
\ee  and $\s_a$, $a=1,2,3$, are the Pauli matrices. The matrices
$t^a$ satisfy the commutation rules
\be  [t^a,~t^b] = 2 i ~\e ^{abc} g_{cd}~ t^d
\ee

We now introduce the operators $a,~a^{\dag}$, which are doublets
under
$SO(2,1)$.  One can show that the generators of $SO(2,1)$ can be
written as
\be  J^a = \ba ~ {t^a \over 2} ~ a
\ee  where $\ba = a^{\dag} \s _3$. The commutation rule for
$a,~a^{\dag}$, compatible with $SO(2,1)$ invariance,  is
\be  [a_i, \ba_j] = \d _{ij},~~~~~~~~~~~~~~i,j=1,2
\ee  We now define the following operators
\beq
 S^a &=& \ba ~t^a t^2 ~\ba^T \cr
 \bS^a= (S^a)^{\dag} & = & a^T ~t^2 t^a ~a
 \label{S}
\eeq   (the superscript $T$ indicates the transpose.) It is easy
to show that both $S$ and
$\bS$ transform as vectors under
$SO(2,1)$ transformations. Further they are null vectors,
\be S^a S_a =0~~,~~~~~~~~~~~~~\bS^a \bS_a =0
\label{nullop}
\ee  and
\be S^a \bS_a = 2(Q^2 -Q)
\label{norm}
\ee where $Q= \sum_{i=1}^{2} \ba_i a_i$. $Q$ is invariant under
$SO(2,1)$ transformations.

The commutation relation between $S$ and $\bS$ is given by
\be [S^a,~\bS^b] = -2 g^{ab} (2Q+2) +8i \e ^{abc} g_{cd} ~J^d
\label{com}
\ee  In showing (\ref{nullop}, \ref{norm}, \ref{com}) the
following properties of the
$t$-matrices were used
\beq  (t^a)_{ij} (t_a)_{kl} &=& - (2 \d _{il} \d_{jk} - \d_{ij}
\d_{kl}) \cr t^a t^b &=& - g^{ab} + i \e ^{abc} g_{cd} t^d
\label{t}
\eeq

Finite dimensional representations of $SO(2,1)$ may be constructed
in terms of Fock states built up using $\ba$ acting on a vacuum
state, with a fixed value of $Q$, say,
$M-1$. A basis of such states is given by $\vert r,s \rangle =
C^{-1}\ba_1^r \ba_2^s
\vert 0\rangle$ with $r+s=M-1$ and $C= \sqrt{r!s!}$. There are $M$
such states and matrix elements of $J^a$ between these states will
give the $(M\times M)$-matrix representation of $SO(2,1)$. We are
interested in the action of $S^a,~ \bS^a$ (or functions of these)
on the states of $\vert r,s \rangle$ of this
$M$-dimensional representation. In this case, we introduce the
rescaled operators
\be
\tilde{S}^a = {S^a \over M}~~,~~~~~~ {\bar{\tilde{S}}}^a = {\bS^a
\over M}
\ee  In the large $M$-limit, as $M \rightarrow \infty$, the
operators $\tilde{S},~
\bar{\tilde{S}}$ commute,
\be [\tilde{S}^a,~ \bar{\tilde{S}} ^b] =0
\label{tildecom}
\ee  Further
\beq && \tilde{S}^a \tilde{S}_a =0~~~,~~~~\bar{\tilde{S}}^a
\bar{\tilde{S}}_a =0
\nonumber \cr && \tilde{S}^a \bar{\tilde{S}}_a =2
\label{*}
\eeq  These properties are just what we have for $n,~{\bar n}$ and
so we can identify
${\tilde S},~{\tilde {\bar S}}$ with $n,{\bar n}$, in the large
$M$-limit.

We are interested in operators ${F}$ made up of equal numbers of
$\tilde{S}$ and
$ \bar{\tilde{S}}$'s. The trace of such an operator over states of
fixed $Q=M-1$ can be written as
\be
\Tr ~{F} = \sum _{r,s=0}^{M-1} {\la r,s| F |r,s \ra}
\label{trace}
\ee   Since $\tilde{S}$'s and $\bar{\tilde{S}}$'s are vectors of
$SO(2,1)$, their traces have to produce invariant tensors of
$SO(2,1)$.  In the large $M$-limit, replacing
${\tilde S},~\bar{{\tilde S}}$ by $n,~{\bar n}$,
$F$ becomes a function of $n,~{\bar n}$ and trace can be
identified as integration. Further, noting that the trace of
identity is $M$, we can define the regularized notion of integrals
of products of
$n,~{\bar n}$ over the Lorentz group as
\be
\left[\int d\mu F(n,~\bn)\right]_{reg} = {1 \over M} \Tr ~F(
\tilde{S},~
\bar{\tilde{S}}) \Biggr] _{M\rightarrow
\infty}
\label{reg}
\ee

As an example of this definition of regularized integrals, we
shall evaluate the integrals of
$n^a \bn^b$ and $n^a n^b \bn^c \bn^d$.  According to our
regularization prescription
\be
\left[ \int d\mu ~n^a \bn^b \right]_{\rm reg} = {1 \over M^3} \Tr
(S^a \bS^b)
\Biggr]_{M
\rightarrow \infty}
\ee   Using the definition of $S^a,~ \bS^b$ in (\ref{S}) and the
properties (\ref{t})  we find that
\be
\left[\int d\mu ~ n^a \bn^b \right]_{\rm reg} = {2 \over 3} ~
g^{ab}
\label{int2}
\ee   The same result can be obtained more efficiently by using
the fact that
$\Tr(S^a \bS^b)$ has to be proportional to the invariant tensor
$g^{ab}$,
\be   {1 \over M^3} \Tr (S^a \bS^b) = x ~ g^{ab}
\ee   The constant of proportionality $x$ is determined by
multiplying both sides by
$g^{ab}$ and using the property (\ref{norm}). Similarly we can
evaluate
\beq
\left[\int d\mu ~n^a n^b \bn^c \bn^d \right]_{\rm reg} & = & {1
\over M^5} \Tr( S^a S^b \bS^c
\bS^d) \Biggr]_{M \rightarrow \infty} \cr & = & -{4 \over 15}
g^{ab} g^{cd} + {4
\over 10} (g^{ac} g^{bd} + g^{ad} g^{bc})\label{int4}
\eeq

The Euclidean integrals corresponding to the above expressions can
be calculated directly and one can verify that their Wick
rotations agree with the above. In other words, we have the result
\be
\left[ {1\over M}{\rm Tr} F({\tilde S}, {\tilde{\bar S}})
\right]_{M\rightarrow \infty}= {\rm Wick ~rotation~ of}~
\left[\int {d\Omega \over 4\pi} F(n, {\bar n})\right]_{Euclidean}
\ee

Given the above procedure of covariantization, we can write down
the covariant version of the mass term (\ref{a18}). We generalize
the derivatives and potentials appearing in
$I(H)$ by defining
$\partial =\half {\tilde S}\cdot \partial, ~\bdel = \half {\tilde
{\bar S}}\cdot \partial$ and
$ A=\half {\tilde S}\cdot A,~ \bA =\half {\tilde{\bar S}}\cdot A$.
The minimal covariant mass term may now be obtained as
\be F_{min}= \int  ~{1\over M} \Tr \left\{ -{2 \pi \over e^2} \S
(H)
\right\} \Biggr]_{M\rightarrow\infty}\label{mass2}
\ee

A final remark on covariantization is that once we have done the
integration over
$(n, {\bar n})$, there will be terms in the action which are
nonlocal in time.  To go back to a Hamiltonian, one needs to
remove this via the use of auxiliary fields, see \cite{VPN2} in
this regard.

\vskip .1in
\noindent$\underline{Covariantized ~expressions}$
\vskip .1in

We now show how the covariantization procedure works specifically
for the mass term. As we show in the appendix, the mass term $F$
in (15) can be written in terms of
$A,~\bA$ and ${\cal A},~\A$ with $\bD{\cal A} - \del \bA=0$, eqs.
(\ref{a7}) to (\ref{a14}). Using the above equation, or eq.
(\ref{a9}), to express $\cal{A}$, $\bar{\cal{A}}$ in terms of
$\bA,~A$ respectively we can write $F$ as \footnote{$F$ may also
be written in terms of the magnetic field $B$ as
\begin{center}
$F={1\over 8e^2} \int \left( \bD^{-1}~B\right)^a \left(
D^{-1}~B\right)^a$
\end{center} }
\be F = {1 \over 2e^2} \int \left( A - \sum_{n=0}^{\infty} (-1)^n
{1 \over \bdel} (\bA {1 \over \bdel})^n \del \bA \right)^a  \left(
\bA - \sum_{n=0}^{\infty}
 (-1)^n {1 \over \del} (A {1 \over \del})^n \bdel A \right)^a
\label{b2}
\ee Let us first consider the term quadratic in $A$'s
\be F^{(2)} = {1 \over 2e^2} \int \left( A^a \bA^a - A^a {1 \over
\del}
\bdel A^a -
\bA^a {1 \over \bdel} \del \bA^a + {1 \over \bdel} (\del \bA^a) {1
\over \del} (\bdel A^a) \right)
\label{b3}
\ee According to the covariantization procedure outlined in
section 3, we get
\beq F^{(2)}_{cov}  &&= \int d\mu ~F^{(2)} \left( A^a \rightarrow
\half {n \cdot A^a}  ,~ \bA^a \rightarrow \half {\bn \cdot A^a} ,~
\del \rightarrow \half {n \cdot \del} ,~ \bdel \rightarrow \half
{\bn \cdot \del}
 \right) \cr &&= {1 \over 8e^2} \int {d^3k \over (2 \pi)^3}
A^a_{\mu} (-k) A^a_{\nu} (k)
\int d\mu \left[ n_{\mu} \bn_{\nu} - n_{\mu}n_{\nu} {{\bn \cdot k}
\over {n \cdot k}} - \bn_{\mu} \bn_{\nu} {{n \cdot k} \over {\bn
\cdot k}} + \bn_{\mu} n_{\nu} \right]
\label{b4}
\eeq The integrals over Lorentz transformations (regularized
expressions) can be evaluated as described in section 3. We have
\beq
 \int d \mu ~ n_{\mu} \bn_{\nu} & = &{2 \over 3}
g_{\mu\nu}~~~~~~~~~~~~~\mu,\nu=1,2,3 \cr
 \int d\mu ~ n_{\mu}n_{\nu} {{\bn \cdot k} \over {n \cdot k}} & =
& -{1 \over 3} g_{\mu\nu} + {{k_{\mu}k_{\nu}} \over k^2}
\label{b5}
\eeq Using (\ref{b5}) in (\ref{b4}) we get
\be F^{(2)}_{cov} = {1 \over 4e^2} \int A^a_{\mu} (-k) \left(
g_{\mu\nu} - {{k_{\mu}k_{\nu}} \over k^2} \right) A^a_{\nu}(k)
\label{b6}
\ee We now consider the term in (\ref{a10}) which is cubic in
$A$'s.
\beq F^{(3)} &&= {1 \over 2e^2} \int f^{abc} \Biggl( \left[ A^a {1
\over \del} (A^b {1
\over \del} \bdel A^c) +  \bA^a {1 \over \bdel}(\bA^b {1
\over \bdel} \del \bA^c)\right] \cr &&~~~~~~~~~~~- \left[ {1 \over
\del} \bdel A^a {1 \over \bdel}(\bA^b {1
\over \bdel} \del \bA^c) + {1 \over \bdel} \del \bA^a {1 \over
\del}(A^b {1
\over \del} \bdel A^c) \right] \Biggr) \cr &&\equiv F^{(3)}_{pure}
+ F^{(3)}_{mixed}
\label{b7}
\eeq where $F^{(3)}_{pure}$ contains only holomorphic or only
antiholomorphic components of $A$'s and $F^{(3)}_{mixed}$ contains
both holomorphic and antiholomorphic components.

According to our covariantization procedure we get in momentum
space,
\beq F^{(3)}_{pure} & =& {i \over 8e^2} \int f^{abc} \d^{(3)}
(p+q+k) A_{\mu}^a (p) A^b_{\nu}(q)
 A^c_{\l}(k) \int d \mu \left( {{\bn
\cdot k} \over {n \cdot p ~ n \cdot k}}n_{\mu}n_{\nu}n_{\l} +
{\rm c.c.}
\right) \cr F^{(3)}_{mixed}& = & {i \over 8e^2} \int f^{abc}
\d^{(3)} (p+q+k) A_{\mu}^a (p)A^b_{\nu}(q)  A^c_{\l} (k)
\int d \mu \left( {{\bn
\cdot p} \over {n \cdot p ~\bn \cdot k}}n_{\mu}n_{\nu}\bn_{\l} +
{\rm c.c.}
\right)
\label{b8}
\eeq where ``c.c" denotes complex conjugation. After
symmetrization over the momenta and integration over the Lorentz
transformations we find
\be F^{(3)}_{pure} = -{i \over 8e^2} \int f^{abc} \d^{(3)}(p+k+q)
 A_{\mu}^a(p) A^b_{\nu}(q) A^c_{\l}(k) ~V_{\mu\nu\l}^{AN} (p,q,k)
 \label{b9}
 \ee
 where
\beq V_{\mu\nu\l}^{AN} (p,q,-(p+q)) &= &{1 \over {p^2q^2-(p \cdot
q)^2}}  \Biggl[~
\left\{{{p \cdot q} \over p^2} - {{q \cdot (q+p)} \over (p+q)^2}
\right\}  p_{\mu} p_{\nu} p_{\l} \cr && + {{p \cdot (p+q)} \over
(p+q)^2} (q_{\mu}q_{\nu}p_{\l} + q_{\l}q_{\nu}p_{\mu} +
q_{\l}q_{\mu}p_{\nu}) -(q \rightarrow p) \Biggr]
\nonumber
\label{b10}
\eeq
$V_{\mu\nu\l}^{AN}$ is proportional to the cubic vertex appearing
in the expression of the magnetic mass proposed by Alexanian and
Nair in \cite{AN}.

The symmetrization over momenta and Lorentz integration is a lot
more involved in the case of $F^{(3)}_{mixed}$ and it was done
using Mathematica. We find that
\be F^{(3)}_{mixed} = {i \over 24e^2} \int  \d^{(3)} (p+k+q)
f^{abc}
 A_{\mu}^a (p) A^b_{\nu}(q) A^c_{\l}(k) \left\{ V_{\mu\nu\l}^{AN}
(p,q,k)
 + L_{\mu\nu\l} (p,q,k) \right\}
 \label{b11}
 \ee where $L_{\mu\nu\l} (p,q,k)$ contains terms involving a
log-dependence on the momenta.

Adding (\ref{b9}) and (\ref{b11}), we find that the total cubic
order contribution  of $F_{cov}$ can be written as
\beq F^{(3)}_{cov} &&=  -{i \over 12e^2} \int f^{abc} \d^{(3)}
(p+k+q)
 A_{\mu}^a (p) A^b_{\nu}(q) A^c_{\l}(k) ~V_{\mu\nu\l}^{AN}
(p,q,k)\nonumber\\
 &&~~~-{{\epsilon_{\mu\nu\lambda}}\over 8e^2} \int \d^{(3)}
(p+k+q) f^{abc}
 \Biggl\{ {2\over q k} \left( {X\over qk-q\cdot k}+{Y\over
qk+q\cdot k}\right)
 {\tilde F}^a_\lambda (p) {\tilde F}^b_\nu (q) {\tilde F}^c_\mu
(k)\nonumber\\
 &&~~~-{2\over qk} \left( {X\over (qk-q\cdot k)^2}-{Y\over
(qk+q\cdot k)^2}\right)
 {\tilde F}^a_\lambda (p) \partial_\nu{\tilde F}^b_\rho (q)
\del_\mu{\tilde F}^c_\rho (k)
 \nonumber\\
 &&~~~-{4\over qk} \left( {X\over (qk-q\cdot k)^2}-{Y\over
(qk+q\cdot k)^2}\right)
 {\tilde F}^a_\lambda (p) \partial_\rho{\tilde F}^b_\nu (q)
\del_\mu{\tilde F}^c_\rho (k)
 \nonumber\\
 &&~~~-{4\over qk} \left( {X\over (qk-q\cdot k)^3}+{Y\over
(qk+q\cdot k)^3}\right)
 {\tilde F}^a_\lambda (p) \partial_\rho\del_\mu{\tilde F}^b_\tau
(q) \del_\nu\del_\tau
 {\tilde F}^c_\rho (k)
 \Biggr\}
\label{b12}
\eeq where $X = {\rm ln} \left[ \left( q k + q \cdot k )/ 2 qk
\right) \right],~  Y = {\rm ln} \left[ \left( q k - q \cdot k) / 2
qk \right)\right] $ and
$\tilde{F}^a_{\mu} = {1\over 2} \e _{\mu\nu\lambda} F_{\nu
\lambda}^a$.

The expression (\ref{b12}) is true up to cubic terms in $A$
although the log-terms were written in terms of
$\tilde{F}^a_{\mu}$ in order to make the gauge invariance more
transparent.

We see that the covariantization of $F$ produces two series of
terms: one series of terms which starts with a term quadratic in
$A$'s and higher order terms necessary for gauge invariance, and a
second series of terms involving the logarithms of momenta which
starts with a term cubic in
$A$'s. These two series of terms are separately gauge invariant.
The non-log terms from  (\ref{b6}) and (\ref{b12}) combine to give
the expression for the magnetic mass term proposed in
\cite{VPN2,AN}. Since this is essentially the covariantization of
$I(H)$, we may conclude that the second series of log-terms
results from the covariantization of just the WZ-term, the term
cubic in
$H^{-1} \del H$, in $I(H)$. After all we have shown in the
appendix, eq. (\ref{a14}), that
\be F(A) = -{2\pi \over e^2} \left[ I(H) - {i
\over {12 \pi}} \int \epsilon ^{\mu \nu \alpha} \Tr ( H^{-1}
\partial _{\mu} H H^{-1}
\partial _{\nu}H H^{-1} \partial _{\alpha}H) \right]
\label{b1}
\ee
\vskip .1in
\noindent{\bf 4. Resummation and magnetic mass}
\vskip .1in

As we have stated earlier, the minimal covariantized mass term in
Euclidean space agrees with  what was proposed in
\cite{VPN2}. The resummation of perturbation theory, to one-loop
order with resummed  propagators and vertices, was carried out in
\cite{AN}. To one-loop order, $\Delta$ was  obtained as $\Delta =
\Delta^{(1)}\approx 1.2(e^2c_A\mu /2\pi)$ The resulting gap
equation $\Delta^{(1)}=\mu^2$ gives a value for the mass gap  as
$\mu \approx 1.2 (e^2 c_A / 2\pi)$. Considering that we are
starting from a perturbative  end with resummation, this value is
quite close to the value $e^2 c_A / 2\pi$  which we found in our
Hamiltonian approach. In the light of all our  discussion above,
this is not so surprising because the mass term used in \cite{AN,
VPN2} has emerged  as the minimal one starting from our
Hamiltonian analysis.  Whether this mass term was anything special
was a question raised by Jackiw and Pi in \cite{JP}. As we have
seen it is a minimal, but not unique, covariant generalization  of
the form which emerges in the Hamiltonian analysis. In the end,
the main advantage of this term might in fact be the following.
Generally nonlocal vertices with covariant Green's functions can
mean that there are additional propagating degrees of freedom in
the theory,  which may be made manifest by checking unitarity via
cutting rules or by making the Lagrangian local via auxiliary
fields. (The Lagrangian then has time-derivatives of the auxiliary
fields which means that they are actually propagating degrees of
freedom.)  For the minimal term, however, the auxiliary fields
have a gauged WZW action  and one can argue that it has no degrees
of freedom modulo the holomorphic symmetry \cite{VPN2}. This
singles out the minimal term to some extent.  Nevertheless, we are
not too far  from what other authors have used. Consider the
nonminimal term $F$ given in (\ref{a15}).  Noting that the field
strength
$B= M^{\dagger -1} {\bar \partial} J M^\dagger = -M \partial {\bar
J} M^{-1}$ and $D^{-1}= M (\partial^{-1}) M^{-1},~ {\bar D}^{-1}=
M^{\dagger -1} ({\bar \partial}^{-1}) M^\dagger$, we see that it
is very similar to, although not exactly, $F_{\mu\nu}({\cal
D}^{-2})F_{\mu\nu}$,  which is the form used by Jackiw and Pi in
\cite{JP}. One could also go further and investigate the gap
equation which results from the use of the covariantized form of
$F$ rather than
$F_{min}$. The additional logarithmic terms in $F$ render the
calculation significantly more  complicated, although there is no
reason to expect the results to be dramatically different.

Now we turn to the question: how do we use this in a calculation?
From a purely  (2+1)-dimensional point of view, we know that there
is no  parameter which controls the resummed loop expansion
\cite{AN,JP}. The calculation of the numerical value  of the gap
in this way would be difficult, at best.  Our Hamiltonian approach
would be better suited to such questions. One can also use the
(2+1)-dimensional theory to describe  magnetic screening in a
quark-gluon plasma in (3+1) dimensions.  Notice that one needs
some perturbative gauge-invariant way of incorporating magnetic
screening for the high temperature calculations with the hard
thermal loop resummations used for  the quark-gluon plasma. More
than specific numerical values, one needs a framework for such
calculations and the present work bears on this issue. (The
embedding of (2+1) results in the (3+1)-dimensional theory has
been discussed in \cite{schulz}.)
\vskip .1in
\noindent{\bf 5. Discussion}
\vskip .1in A number of different concepts have been brought
together in this work and  it may be useful to summarize briefly
what we have done. Based on our Hamiltonian analysis, one can show
that there is a mass term of the form $(\del \vf^a \bdel \vf^a)$
at the lowest nontrivial order in $\vf^a$.  There is no ambiguity
to this order in
$\vf^a$. In generalizing from this, first of all, we need to write
down an expression in terms of $H=e^{t^a\vf^a}$ which reduces to
$(\del \vf^a
\bdel \vf^a)$ at the lowest order. There are many such expressions.
In the appendix, we outline the reasons why $(\del
\vf^a \bdel \vf^a)$ should be considered as the lowest order term
of $\Tr (\del H \bdel H^{-1})$. The argument then is to use this
term, or some generalization of it, as a mass term to be used in a
resummation procedure. Even at this stage, although some
restrictions on the possible form of a mass term have been
obtained, there are still many terms which have holomorphic
invariance and agree with
$\Tr (\del H \bdel H^{-1})$ to the requisite order, $F_{min}$ in
Eq.(\ref{a18}) and $F$ in Eq.(\ref{a15}) being two such
expressions.
$F_{min}$ is a minimal one in the sense of not requiring
additional use of Green's functions and, for this reason, leads to
simpler formulae upon covariantization.

Once we have chosen a specific mass term such as $F_{min}$, its
use in an action formalism, rather than in a Hamiltonian analysis,
will require that it be covariantized to maintain Lorentz
covariance order by order. We have given a method of
covariantization, both in Minkowski space and in the Wick rotated
Euclidean case. Finally, we have given a discussion of the results
of the resummation carried out with the minimal mass term
$F_{min}$.
\vskip .1in
\noindent{\bf APPENDIX}
\vskip .1in
\def\theequation{A\arabic{equation}}
\reset We have written the mass term in (\ref{10}) to the second
order in $\vf$.  We want to write an expression in terms of $H$
for which this is  the quadratic expansion and show that the
correct expression should be
$\Tr (\partial H \bdel H^{-1})$ and not something like
$\Tr (\partial H \bdel H)$.

Writing the kinetic energy $T$ as
\be T = -{e^2 \over 2} {\d^2 \over{\d A_i^a \d A_i^a}}
\label {a1}
\ee we have
\beq && {e^2 \over 2} \int d\mu (H) e^{2 c_A \S} \Psi^{*}
\left(-{\d^2 \over{\d A_i^a
\d A_i^a}}
\Psi\right) \cr &&= {e^2 \over 2} \int d\mu (H) \Phi^* \left[
-{{\d^2 \Phi} \over{\d A_i^a
\d A_i^a}} +2 c_A {\d \S \over \d A_i^a} {\d \Phi \over \d A_i^a}
- \left(c_A^2 {\d \S \over
\d A_i^a} {\d \S
\over \d A_i^a} - c_A {{\d^2 \S} \over{\d A_i^a \d A_i^a}} \right)
\Phi \right]
\label{a2}
\eeq where we have written $\Psi = e^{-c_A \S} \Phi$, absorbing
the crucial WZW-part of the measure into the wavefunctions. In a
similar way we have
\beq && {e^2 \over 2} \int d\mu (H) e^{2 c_A \S} \left(-{\d^2
\over{\d A_i^a \d A_i^a}}
\Psi^*\right) \Psi \cr &&= {e^2 \over 2} \int d\mu (H)  \left[
-{{\d^2 \Phi^*} \over{\d A_i^a \d A_i^a}} +2 c_A {\d \S \over \d
A_i^a} {\d \Phi^* \over \d A_i^a} - \left(c_A^2 {\d \S
\over \d A_i^a} {\d
\S
\over \d A_i^a} - c_A {{\d^2 \S} \over{\d A_i^a \d A_i^a}} \right)
\Phi^* \right] \Phi
\label{a3}
\eeq We now add these two equations and do a partial integration
for $\d /\d A_i^a$.  In doing so we have to use the full measure
$d\mu(H) e^{2c_A \S} = [dA d\bA]/({\rm vol} \G)$. This gives
\beq
\int d\mu(H) {\d \S \over \d A_i^a} {{\d (\Phi^* \Phi)} \over \d
A_i^a}  & = & \int d\mu(H) e^{2 c_A \S} e^{-2 c_A \S}  {\d \S
\over \d A_i^a}{{\d (\Phi^* \Phi)}
\over \d A_i^a} \cr & = & \int d\mu(H) \left(-{\d^2 \S \over {\d
A_i^a \d A_i^a}} + 2 c_A  {\d \S
\over \d A_i^a}{\d \S
\over \d A_i^a}\right)\Phi^* \Phi
\label{a4}
\eeq Thus upon adding (\ref{a2}) and (\ref{a3}) and using
(\ref{a4}) we find
\be
\la \Psi | T | \Psi \ra = \half \left\la \Phi \left| {{\tilde{T} +
\tilde{T} ^{\dag} }} \right|
\Phi
\right\ra + {{e^2 c_A^2} \over 2}  \left\la \Phi \left| {\d \S
\over \d A_i^a} {\d \S \over \d A_i^a}
\right|
\Phi
\right\ra
\label{a5}
\ee where the inner product in terms of $\Phi$'s is now
\be
\la 1 | 2 \ra =\int d\mu(H) \Phi^* \Phi
\label {a6}
\ee and $\tilde{T} \Phi = - {e^2 \over 2} {{\d ^2 \Phi} \over {\d
A_i \d A_i}}$.
$\tilde{T} ^{\dag}$ denotes the adjoint of $\tilde{T}$ with just
the Haar measure for integration as in (\ref{a6}). Eq.(\ref{a5})
displays the extra ``mass term" as
\be {{e^2 c_A^2} \over 2}  \int {\d \S \over \d A_i^a} {\d \S
\over \d A_i^a} =  {{e^2 c_A^2} \over 2}
\int {\d \S \over
\d A^a} {\d \S \over \d \bA^a} \equiv m^2 F
\label{a7}
\ee where $m= e^2 c_A /2 \pi$. In terms of the gauge potentials,
the lowest order term of this expression, viz., the quadratic term,
is the mass term (\ref{c1}) for $A$'s, the higher order terms
being required for reasons of gauge invariance. This term can be
simplified further as follows. Regarding
$\S$ as a function of $A,~\bA$, we can write its variation as
\be
\d \S = -{1 \over 2\pi} \int (A- {\cal A})^a \d \bA^a + (\bA -
\bar{\cal A})^a
 \d A^a
\label{a8}
\ee where $\cal{A}$, $\bar{\cal A}$ obey the equations
\beq && \bar{D} {\cal A} - \del \bA =0 \cr && D \bar{\cal A} -
\bdel A =0
\label{a9}
\eeq This shows that we may write
\be F ={{2 \pi^2} \over e^2} \int {\d \S \over
\d A^a} {\d \S \over \d \bA^a} = {1 \over 2e^2} \int (A- {\cal
A})^a  (\bA - \bar{\cal A})^a
\label{a10}
\ee Taking the variation of (\ref{a10}) and using (\ref{a8}) we
find
\be -{e^2 \over  \pi} \d F =  \d \S (A, \bA) - { 1 \over 2 \pi}
\int ({\cal A}-
 A)^a
\d \bar{{\cal A}}^a + (\bar{{\cal A}} - \bA)^a \d {\cal A}^a
\label{a11}
\ee Notice that the second term is just like the variation of $\S$
as in (\ref{a8}), except for the exchange ${\cal A} \rightarrow A,
~\bar{{\cal A}} \rightarrow \bar{A}$. In terms of the
parametrization (1) of $A,~\bA$, we can solve (\ref{a9}) to get
\beq && \bar{{\cal A}}= - \bdel M M^{-1} =M \bdel M^{-1} \cr &&
{\cal A}=  M^{\dag -1}\del M^{\dag} = - \del M^{\dag -1} M^{\dag}
\label{a12}
\eeq The exchange ${\cal A} \rightarrow A, ~ \bar{\cal A}
\rightarrow \bA$ thus corresponds to $M \rightarrow M^{\dag -1}$
or $H= M^{\dag} M \rightarrow H^{-1}= M^{-1} M^{\dag -1}$.
Equation (\ref{a11}) can thus be written as
\be -{e^2 \over  \pi} \d F =  \d \S (H) +  \d \S (H^{-1})
\label{a13}
\ee This implies
\be F =-{ \pi \over e^2} \left[ \S (H) + \S (H^{-1}) \right] = -{1
\over e^2} \Tr (\bdel H
\del H^{-1})
\label{a14}
\ee

This brings us to the point of identifying the mass term which
satisfies the requirements 1 and 2 listed in section 2, but not
yet the requirement 3. The expression for $F$ as it is written in
(\ref{a14}) is not holomorphically invariant. This is because,
eventhough $I(H)$ is invariant, $\S (H^{-1})$ is not. (Notice that
the inversion of $D,~\bD$ to obtain ${\cal A},~\A$, or
equivalently the solution (\ref{a12}), requires fixing a
``holomorphic frame". This is why the form of $F$ in (\ref{a14})
is not holomorphically invariant.)  A holomorphically invariant
completion of
$F$ is straightforward. Notice that $F$ in (\ref{a14}) is
proportional to the kinetic term of $\S (H)$. Since $\S (H)$ is
invariant under $H \rightarrow V H \bV$, we see that a minimal
completion of $F$ we can use is
\be F \rightarrow F_{min}= -{2 \pi \over e^2} \S (H)
\label{}
\ee The minimal mass term is then the WZW action.

\vskip .1in
\noindent{\bf Acknowledgements}
\vskip .1in This work was supported in part by the National
Science Foundation grants PHY-9970724 and PHY-9605216 and the
PSC-CUNY-30 awards. CK thanks Lehman College of CUNY and
Rockefeller University for hospitality facilitating the completion
of this work.

\end{document}